\documentclass[a4paper,11pt]{article}
\pdfoutput=1
\usepackage{pos}

\title{Proton and neutron electromagnetic radii and magnetic moments from \texorpdfstring{$N_f = 2 + 1$}{Nf=2+1} lattice QCD}
\ShortTitle{Proton and neutron electromagnetic radii and magnetic moments from lattice QCD}

\author*[a]{Miguel Salg}
\author[b,c]{Dalibor Djukanovic}
\author[a]{Georg von Hippel}
\author[a,b]{Harvey B. Meyer}
\author[a]{Konstantin Ottnad}
\author[a,b]{Hartmut Wittig}

\affiliation[a]{\texorpdfstring{PRISMA${}^+$}{PRISMA+} Cluster of Excellence and Institute for Nuclear Physics, Johannes Gutenberg University Mainz, Johann-Joachim-Becher-Weg 45, 55128 Mainz, Germany}
\affiliation[b]{Helmholtz Institute Mainz, Staudingerweg 18, 55128 Mainz, Germany}
\affiliation[c]{GSI Helmholtzzentrum für Schwerionenforschung, 64291 Darmstadt, Germany}

\emailAdd{msalg@uni-mainz.de}

\abstract{
We present results for the electromagnetic form factors of the proton and neutron computed on the $(2 + 1)$-flavor Coordinated Lattice Simulations (CLS) ensembles including both quark-connected and -disconnected contributions.
The $Q^2$-, pion-mass, lattice-spacing, and finite-volume dependence of our form factor data is fitted simultaneously to the expressions resulting from covariant chiral perturbation theory including vector mesons amended by models for lattice artefacts.
From these fits, we determine the electric and magnetic radii and the magnetic moments of the proton and neutron, as well as the Zemach radius of the proton.
To assess the influence of systematic effects, we average over various cuts in the pion mass and the momentum transfer, as well as over different models for the lattice-spacing and finite-volume dependence, using weights derived from the Akaike Information Criterion (AIC).
Our results for the magnetic moments of the proton and neutron are in good agreement with the experimental values and have a relative precision of about \qty{2.4}{\percent} and \qty{3.7}{\percent}, respectively.
For the electromagnetic radii of the proton, we achieve a precision at the \qty{1.5}{\percent} level. 
\vspace{0.5cm}
\begin{flushright}
    MITP-23-077
\end{flushright}
}

\FullConference{The 40th International Symposium on Lattice Field Theory (Lattice 2023)\\
July 31st - August 4th, 2023\\
Fermi National Accelerator Laboratory\\}

\makeatletter
\hypersetup{pdftitle=\@title, pdfauthor=\printHeadAuthors}
\makeatother
\usepackage[USenglish]{babel}
\usepackage[group-digits=integer, exponent-product= \cdot]{siunitx}
\usepackage{csquotes}
\usepackage[english]{cleveref}
\usepackage[all]{foreign}
\usepackage{booktabs}
\usepackage{tikz}
\usepackage{braket}
\DeclareMathOperator{\tr}{tr}

\begin{document}
\maketitle

\section{Introduction}
The so-called \enquote{proton radius puzzle}, \ie the tension between different measurements of the proton's electric radius, has gripped the scientific community for more than 10 years \cite{Karr2020}.
While most of the recent experiments point towards a smaller electric radius, so that this puzzle is approaching its resolution, the situation regarding the magnetic radius is still less clear, as one also finds discrepant results for this quantity \cite{Lee2015}.

In lattice QCD as in scattering experiments, the radii are extracted from the slope of the electromagnetic form factors at $Q^2 = 0$.
A full theoretical prediction of the latter necessitates the calculation of quark-disconnected diagrams, which are computationally very expensive and intrinsically very noisy.
Therefore, they have been neglected in most previous lattice studies.
In particular, our calculation is the first to simultaneously evaluate all contributions and extrapolate to the continuum and infinite-volume limits.
This presentation is based on Refs.\@ \cite{Djukanovic2023,Djukanovic2023a,Djukanovic2023b}, to which we refer the interested reader for more details on our computational setup as well as on our analysis.

\section{Lattice setup}
We use a set of ensembles which have been generated by CLS \cite{Bruno2015} with $2 + 1$ flavors of non-perturbatively $\mathcal{O}(a)$-improved Wilson fermions \cite{Sheikholeslami1985,Bulava2013} and a tree-level improved Lüscher-Weisz gauge action \cite{Luescher1985}.
All the ensembles we employ follow the chiral trajectory characterized by $\tr M_q = 2m_l + m_s = \text{const}$.
\Cref{tab:ensembles} displays the set of ensembles entering the analysis:
they cover four lattice spacings in the range from \qty{0.050}{fm} to \qty{0.086}{fm}, and several different pion masses, including one slightly below the physical value (E250).

\begin{table}[htb]
    \begin{minipage}{\textwidth}
        \centering
        \renewcommand*{\thefootnote}{\alph{footnote}}
        \renewcommand*\footnoterule{}
        \begin{tabular}{lcccccccc}
            \toprule
            ID                   & $\beta$ & $t_0^\mathrm{sym}/a^2$ & $T/a$ & $L/a$ & $M_\pi$ [MeV] & $N_\mathrm{cfg}^\mathrm{conn}$ & $N_\mathrm{cfg}^\mathrm{disc}$ & $t_\mathrm{sep}/a$ \\ \midrule
            C101                 & 3.40    & 2.860(11)              & 96    & 48    & 227           & 1988                           & 994                            & 4 -- 17            \\
            N101\footnotemark[1] & 3.40    & 2.860(11)              & 128   & 48    & 283           & 1588                           & 1588                           & 4 -- 17            \\
            H105\footnotemark[1] & 3.40    & 2.860(11)              & 96    & 32    & 283           & 1024                           & 1024                           & 4 -- 17            \\[\defaultaddspace]
            D450                 & 3.46    & 3.659(16)              & 128   & 64    & 218           & 498                            & 498                            & 4 -- 20            \\
            N451\footnotemark[1] & 3.46    & 3.659(16)              & 128   & 48    & 289           & 1010                           & 1010                           & 4 -- 20 (stride 2) \\[\defaultaddspace]
            E250                 & 3.55    & 5.164(18)              & 192   & 96    & 130           & 398                            & 796                            & 4 -- 22 (stride 2) \\
            D200                 & 3.55    & 5.164(18)              & 128   & 64    & 207           & 1996                           & 998                            & 4 -- 22 (stride 2) \\
            N200\footnotemark[1] & 3.55    & 5.164(18)              & 128   & 48    & 281           & 1708                           & 1708                           & 4 -- 22 (stride 2) \\
            S201\footnotemark[1] & 3.55    & 5.164(18)              & 128   & 32    & 295           & 2092                           & 2092                           & 4 -- 22 (stride 2) \\[\defaultaddspace]
            E300                 & 3.70    & 8.595(29)              & 192   & 96    & 176           & 569                            & 569                            & 4 -- 28 (stride 2) \\
            J303                 & 3.70    & 8.595(29)              & 192   & 64    & 266           & 1073                           & 1073                           & 4 -- 28 (stride 2) \\ \bottomrule
        \end{tabular}
        \footnotetext[1]{These ensembles are not used in the final fits but only to constrain discretization and finite-volume effects.}
    \end{minipage}
    \caption{Overview of the ensembles used in this study. Further details are contained in table~I of Ref.\@ \cite{Djukanovic2023}.}
    \label{tab:ensembles}
\end{table}

On these ensembles, we measure the two- and three-point correlation functions of the nucleon.
For the three-point functions, the pertinent Wick contractions yield a connected and a disconnected contribution.
The disconnected part is constructed from the quark loops and the two-point functions, where the former are computed via stochastic estimation using a frequency-splitting technique \cite{Giusti2019} and the one-end trick \cite{McNeile2006}.
Our procedure is described in detail in Ref.\@ \cite{Ce2022}.
We use a symmetrized and $\mathcal{O}(a)$-improved conserved vector current \cite{Gerardin2019a}, so that no renormalization is required.
From the two- and three-point correlation functions, we extract the effective form factors in the isospin basis using the ratio method \cite{Korzec2009} and the same estimators for the effective electric and magnetic form factors as in Ref.\@ \cite{Djukanovic2021}.
We express all dimensionful quantities in units of $t_0$ using the determination of $t_0^\mathrm{sym}/a^2$ from Ref.\@ \cite{Bruno2017}.
Only our final results for the radii are converted to physical units by means of the FLAG estimate \cite{Aoki2021} $\sqrt{t_{0, \mathrm{phys}}} = \qty{0.14464(87)}{fm}$ for $N_f = 2 + 1$.

\begin{figure}[htb]
    \begin{center}
       \includegraphics[width=0.95\textwidth]{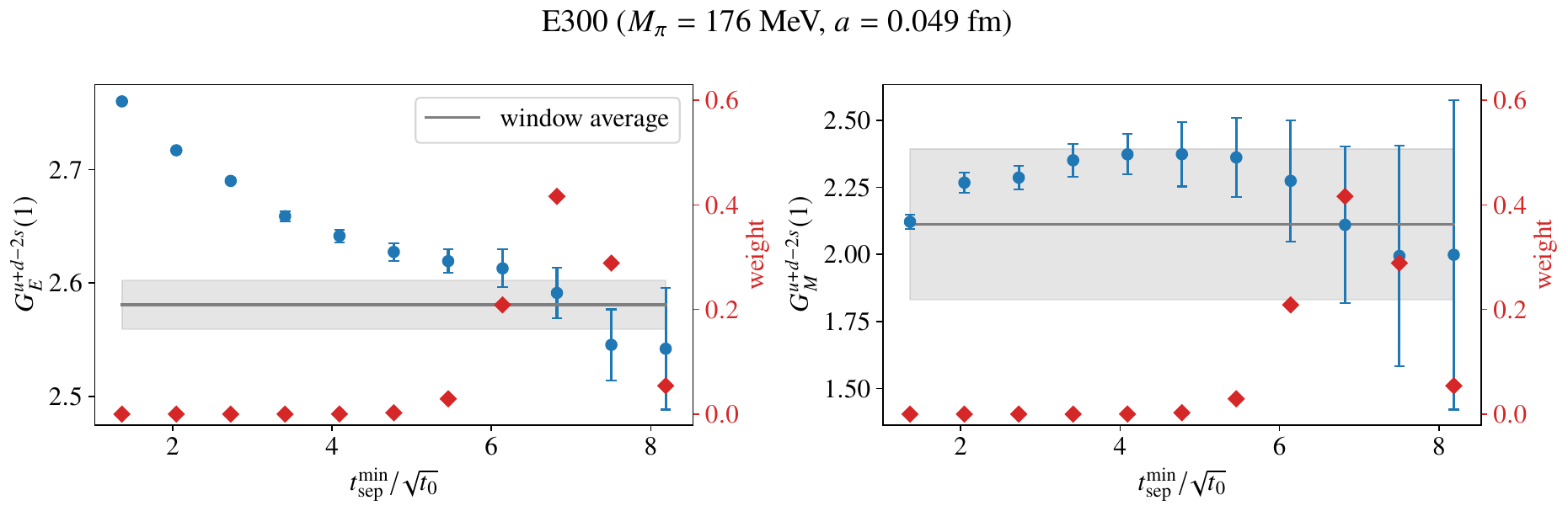}
    \end{center}
    \caption{Isoscalar electromagnetic form factors at the first non-vanishing momentum ($Q^2 \approx \qty{0.067}{GeV^2}$) on the ensemble E300 as a function of the minimal source-sink separation entering the summation fit. Each blue point corresponds to a single fit starting at the value given on the horizontal axis. The associated weights are represented by the red diamonds, with the gray curves and bands depicting the averaged results.}
    \label{fig:window_average}
\end{figure}

In order to treat the excited-state systematics, we employ the summation method \cite{Capitani2012}, where we vary the starting values $t_\mathrm{sep}^\mathrm{min}$ of the linear fits.
In the next step, we perform a weighted average over $t_\mathrm{sep}^\mathrm{min}$, where the weights are given by a smooth window function \cite{Djukanovic2022,Agadjanov2023}.
This averaging strategy is illustrated in \cref{fig:window_average} for the isoscalar combination at the first non-vanishing momentum on E300.
For further details, we refer to section~III of Ref.\@ \cite{Djukanovic2023}.

\section{Direct Baryon \texorpdfstring{$\chi$PT}{ChPT} fits}
Since the radii are defined in terms of the $Q^2$-dependence of the form factors, a parametrization of the latter is required.
We combine this with the extrapolation to the physical point ($M_\pi = M_{\pi, \mathrm{phys}}$, $a = 0$, $L = \infty$) by performing a simultaneous fit of the $Q^2$-, pion-mass, lattice-spacing, and finite-volume dependence of the form factors to the expressions resulting from covariant baryon chiral perturbation theory (B$\chi$PT) \cite{Bauer2012}.
While explicit $\Delta$ degrees of freedom are not considered in the fit, we include the contributions of the relevant vector mesons, \ie $\rho$ in the isovector channel and $\omega$ and $\phi$ in the isoscalar channel.
In this way, we can extend the validity of the expressions up to $Q^2 \lesssim M_\rho^2 \approx \qty{0.6}{GeV^2}$ \cite{Kubis2001,Bauer2012}.
Performing the fits for $G_E$ and $G_M$ simultaneously allows us to treat the correlations not only between different $Q^2$, but also between $G_E$ and $G_M$ correctly.
The physical pion mass $M_{\pi, \mathrm{phys}}$ is fixed in units of $\sqrt{t_0}$ using its value in the isospin limit \cite{Aoki2014}.

We perform several such fits with various cuts in the pion mass ($M_\pi \leq \qty{0.23}{GeV}$ and $M_\pi \leq \qty{0.27}{GeV}$) and the momentum transfer ($Q^2 \leq \qtyrange[range-phrase = {, \ldots, }, range-units = single]{0.3}{0.6}{GeV^2}$), as well as with different models for the lattice-spacing and/or finite-volume dependence.
Finally, we reconstruct the proton and neutron form factors as linear combinations of the B$\chi$PT formulae for the isovector and isoscalar channels, evaluating the low-energy constants as determined from the separate fits in these channels.

One major benefit of this method compared to the more traditional approach of fitting the $Q^2$-dependence on each ensemble individually and afterwards extrapolating to the physical point is the following: performing direct fits leads to a much larger number of degrees of freedom entering the fit, which increases the stability against lowering the applied momentum cut considerably.
The inclusion of several ensembles in one fit also decreases the errors on the resulting radii significantly.

\section{Zemach radius of the proton}
Our results for the electromagnetic form factors can be used to compute, in addition to the electric and magnetic radii, the Zemach radius of the proton,
\begin{equation}
    r_Z^p = -\frac{4}{\pi} \int_0^\infty \frac{dQ}{Q^2} \left[\frac{G_E^p(Q^2) G_M^p(Q^2)}{\mu_M^p} - 1\right] = -\frac{2}{\pi} \int_0^\infty \frac{dQ^2}{(Q^2)^{3/2}} \left[\frac{G_E^p(Q^2) G_M^p(Q^2)}{\mu_M^p} - 1\right] ,
    \label{eq:Zemach_radius}
\end{equation}
which determines the leading-order proton-structure contribution to the $S$-state hyperfine splitting (HFS) of hydrogen \cite{Zemach1956}.
A firm theoretical prediction of the Zemach radius is of crucial importance for the next generation of atomic spectroscopy experiments on muonic hydrogen \cite{Sato2014,Pizzolotto2020,Amaro2022}.

Due to their very limited range of validity in $Q^2$, the B$\chi$PT fits cannot be employed directly to evaluate the full integral in \cref{eq:Zemach_radius}.
Therefore, we extrapolate the results for $G_E^p$ and $G_M^p$ from each variation of the B$\chi$PT fits using a model-independent \ansatz based on the $z$-expansion \cite{Hill2010}.
We incorporate the four sum rules from Ref.\@ \cite{Lee2015} for each form factor, which ensure the correct asymptotic behavior of the latter for large $Q^2$ \cite{Lepage1980}.
For the numerical integration of \cref{eq:Zemach_radius}, we smoothly replace the B$\chi$PT parametrization of the form factors by the $z$-expansion-based extrapolation in a narrow window around the $Q^2$-cut of the corresponding model variation.

Because of its strong fall-off with $Q^2$, the form-factor term in \cref{eq:Zemach_radius} at $Q^2 > \qty{0.6}{GeV^2}$ contributes less than \qty{0.9}{\percent} to the Zemach radius of the proton.
Hence, the contribution of the extrapolated form factors is highly suppressed, so that the precise form of the chosen model for the extrapolation only has a marginal influence on our result for $r_Z^p$.
Finally, we note that the major advantage of our approach based on the B$\chi$PT fits over an integration of the form factors on each ensemble is that the Zemach radius can be computed directly at the physical point.

\section{Model average and final results}
Since we do not have a strong \apriori preference for one specific setup of the B$\chi$PT fits, we determine our final results and total errors from model averages over different fit variations.
For this purpose, we use weights derived from the Akaike Information Criterion \cite{Akaike1974,Neil2022}.
To estimate the statistical and systematic errors of our model averages, we adopt a bootstrapped variant of the method from Ref.\@ \cite{Borsanyi2021}.
Our final results are collected in \cref{tab:final_results}.
We find that we can obtain the magnetic radii of the proton and neutron to a precision very similar to their respective electric radii.

\begin{table}[htb]
    \centering
    \begin{tabular}{llllc}
        \toprule
        Channel   & \multicolumn{1}{c}{$\langle r_E^2 \rangle$ [\unit{fm^2}]} & \multicolumn{1}{c}{$\langle r_M^2 \rangle$ [\unit{fm^2}]} & \multicolumn{1}{c}{$\mu_M$} & $r_Z$ [fm]      \\ \midrule
        Isovector & $0.785(22)(26)$                                           & $0.663(11)(8)$                                            & $4.62(10)(7)$               & --              \\
        Isoscalar & $0.554(18)(13)$                                           & $0.657(30)(31)$                                           & $2.47(11)(10)$              & --              \\
        Proton    & $0.672(14)(18)$                                           & $0.658(12)(8)$                                            & $2.739(63)(18)$             & $1.013(10)(12)$ \\
        Neutron   & $-0.115(13)(7)$                                           & $0.667(11)(16)$                                           & $-1.893(39)(58)$            & --              \\ \bottomrule
    \end{tabular}
    \caption{Final results for the radii and magnetic moments. In each case, the first error is statistical and the second one systematic, respectively.}
    \label{tab:final_results}
\end{table}

\begin{figure}[htb]
    \begin{center}
        \includegraphics[width=0.95\textwidth]{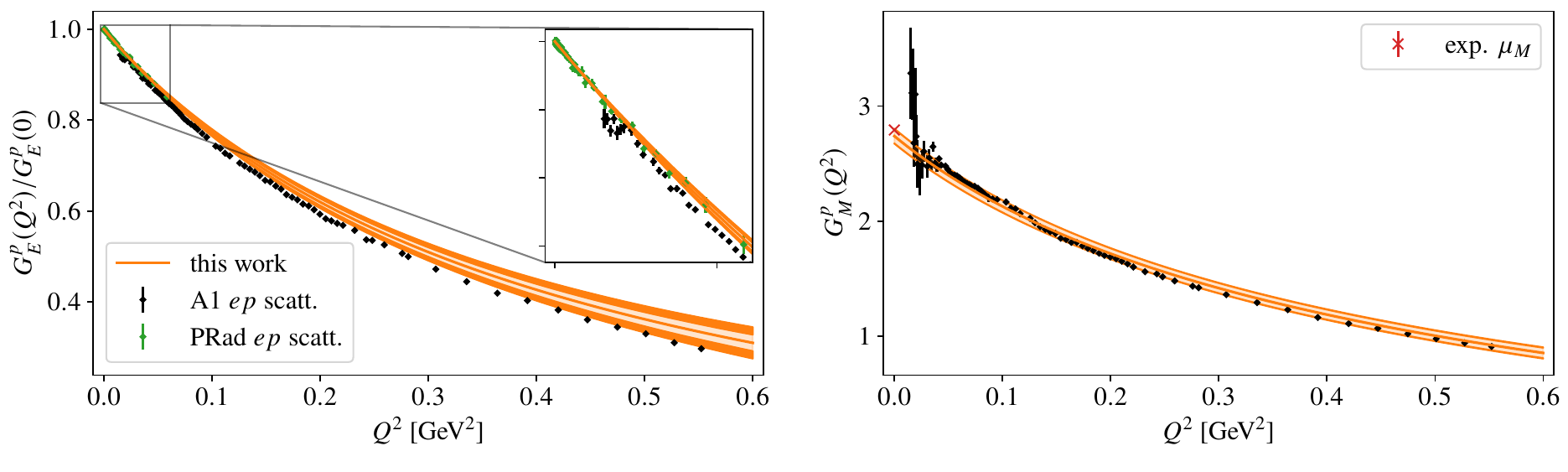}
        \includegraphics[width=0.95\textwidth]{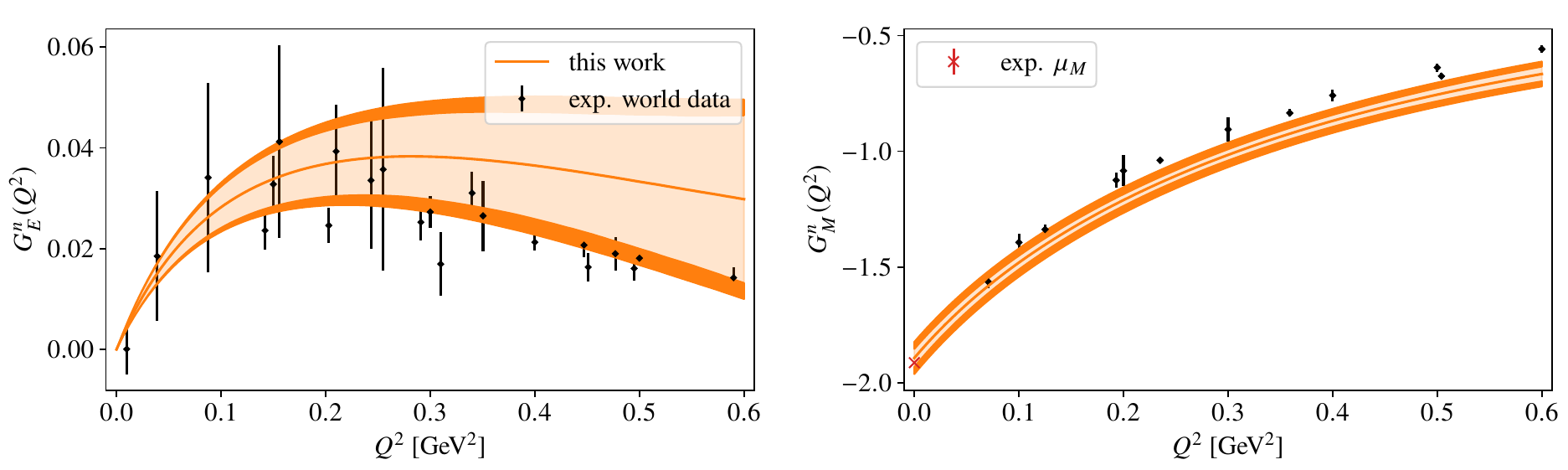}
    \end{center}
    \caption{Electromagnetic form factors of the proton and neutron at the physical point as a function of $Q^2$.
      The orange curves and light (dark) orange bands correspond to our final results with their statistical (full) uncertainties.
      For the proton, the black diamonds represent the experimental $ep$-scattering data from Mainz/A1 \cite{Bernauer2014} obtained using Rosenbluth separation, while the green diamonds represent the data from PRad \cite{Xiong2019}.
      For the neutron, the black diamonds show the experimental world data collected in Ref.\@ \cite{Ye2018}.
      The experimental values of the magnetic moments \cite{Workman2022} are depicted by red crosses.}
    \label{fig:bchpt_fits_model_average}
\end{figure}

To further compare our results to experiment we perform model averages of the form factors themselves.
These are plotted in \cref{fig:bchpt_fits_model_average} for the proton and neutron.
One observes that our slope of $G_E^p$ is much closer to that of PRad \cite{Xiong2019} than to that of A1 \cite{Bernauer2014}, while $G_M^p$ agrees well with A1.
For the neutron, we compare with the collected experimental world data \cite{Ye2018}, which are largely compatible with our curves within our quoted errors.
Furthermore, our results reproduce within their errors the experimental values of the magnetic moments of the proton and of the neutron \cite{Workman2022}.

\begin{figure}[htb]
    \begin{center}
        \includegraphics[width=0.95\textwidth]{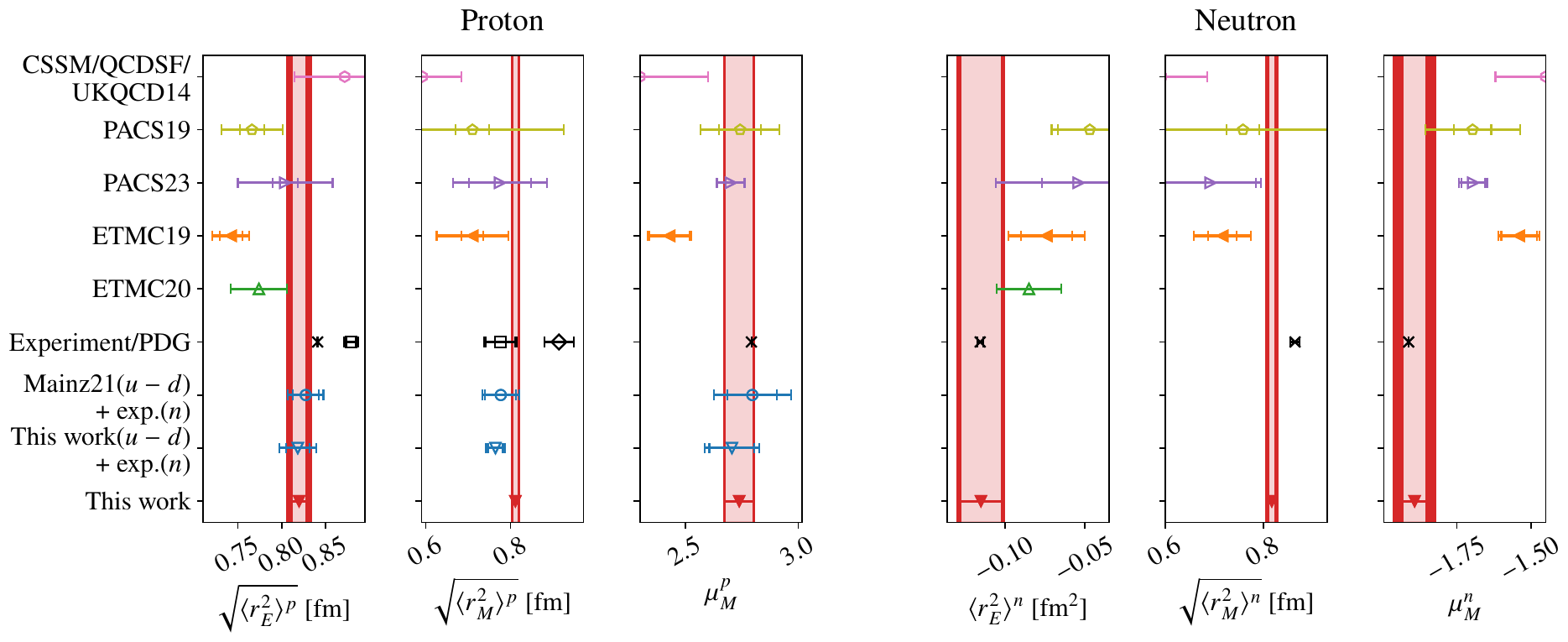}
    \end{center}
    \caption{Comparison of our best estimates for the electromagnetic radii and the magnetic moments of the proton and neutron with other lattice calculations \cite{Djukanovic2021,Alexandrou2020,Alexandrou2019,Tsuji2023,Shintani2019,Shanahan2014a,Shanahan2014}. The experimental values for the neutron and for $\mu_M^p$ are taken from PDG \cite{Workman2022}. The two data points for $r_E^p$ depict the values from PDG \cite{Workman2022} (cross) and Mainz/A1 \cite{Bernauer2014} (square), respectively. For $r_M^p$, on the other hand, they show the reanalysis of Ref.\@ \cite{Lee2015} either using the world data excluding that of Ref.\@ \cite{Bernauer2014} (diamond) or using only that of Ref.\@ \cite{Bernauer2014} (square).}
    \label{fig:comparison_em_radii}
\end{figure}

In \cref{fig:comparison_em_radii}, our results for the electromagnetic radii and magnetic moments of the proton and neutron are compared to recent lattice determinations \cite{Alexandrou2020,Alexandrou2019,Tsuji2023,Shintani2019,Shanahan2014a,Shanahan2014} and to the experimental values.
We remark that the only other lattice study including disconnected diagrams is ETMC19 \cite{Alexandrou2019}, which, however, does not perform a continuum and infinite-volume extrapolation.
By and large, we observe a reasonable agreement with other lattice determinations, where our results are in general closer to the experimental values than those of ETMC19, in particular for the magnetic moments.
Our error estimates for the electric radii and magnetic moments are comparable with the other lattice studies, while being substantially smaller for the magnetic radii, which is due to our direct fit approach.

As is the case for most of the other recent lattice calculations \cite{Alexandrou2020,Alexandrou2019,Tsuji2023,Shintani2019}, our result for $r_E^p$ is much closer to the PDG value \cite{Workman2022}, which is completely dominated by muonic hydrogen spectroscopy, and to the result of the PRad $ep$-scattering experiment \cite{Xiong2019} than to the A1 $ep$-scattering result \cite{Bernauer2014}.
For $r_M^p$, on the other hand, our estimate is well compatible with the value inferred from the A1 experiment by the analyses \cite{Bernauer2014,Lee2015} and is in some tension with the other collected world data \cite{Lee2015}.
As can be seen from \cref{fig:bchpt_fits_model_average} (top right), the good agreement with A1 is not only observed in the magnetic radius, but also for the magnetic form factor over the whole range of $Q^2$ under study.
We note that the dispersive analysis of the Mainz/A1 and PRad data \cite{Lin2021a} has yielded a significantly larger magnetic radius than the $z$-expansion-based analysis of the Mainz/A1 data \cite{Lee2015}.

\begin{figure}[htb]
    \begin{center}
        \includegraphics[width=0.5\textwidth]{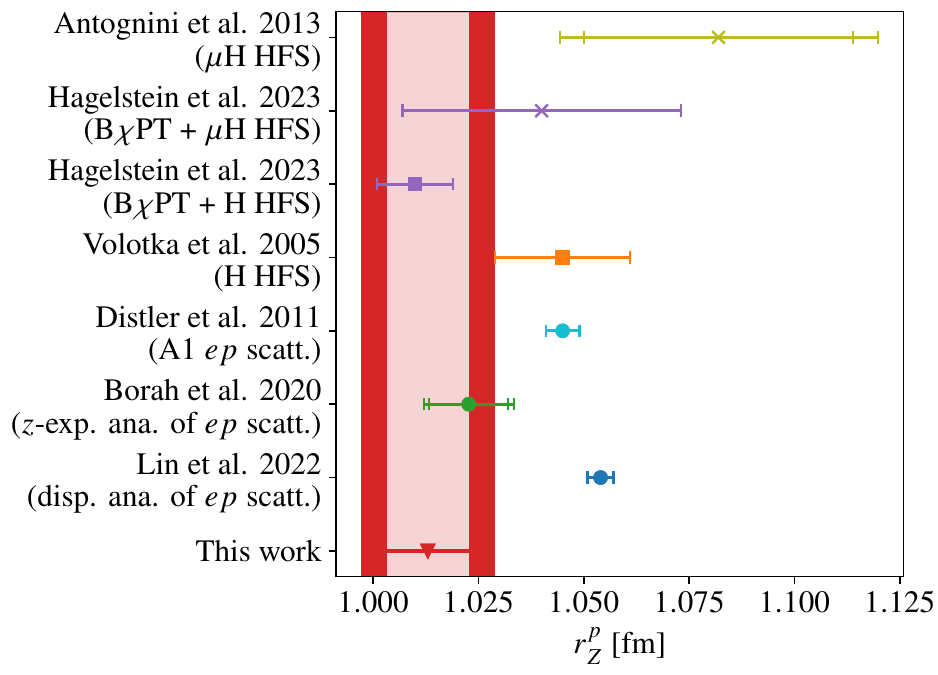}
    \end{center}
    \caption{Comparison of our best estimate for the Zemach radius of the proton with determinations based on experimental data, \ie muonic hydrogen HFS \cite{Antognini2013,Hagelstein2023} (crosses), electronic hydrogen HFS \cite{Volotka2005,Hagelstein2023} (squares), and $ep$ scattering \cite{Distler2011,Borah2020,Lin2022} (circles).}
    \label{fig:comparison_Zemach_radius}
\end{figure}

In \cref{fig:comparison_Zemach_radius}, our result for the Zemach radius of the proton is compared to other determinations based on experimental data \cite{Antognini2013,Hagelstein2023,Volotka2005,Distler2011,Borah2020,Lin2022}.
While our result is compatible with most of these extractions, we observe a tension with the dispersive analysis of $ep$-scattering data \cite{Lin2022}.
We also note that our estimate is smaller than almost all of the experimental determinations.
In interpreting our result for the Zemach radius, one must take into account that it is not independent from that for the electromagnetic radii because it is based on the same lattice data for the form factors and the same B$\chi$PT fits.
Hence, our small results for $r_E^p$ and $r_M^p$ naturally imply a small value for $r_Z^p$.

The aforementioned relatively large result for the magnetic proton radius from dispersive analyses \cite{Lin2021a,Lin2022} may explain why we observe a tension in the Zemach radius with Ref.\@ \cite{Lin2022}.
For a deeper understanding of the underlying differences, a comparison of the full $Q^2$-dependence of the form factors would be required, rather than merely of the radii.

\section{Conclusions}
In these proceedings, we have investigated the electromagnetic form factors of the proton and neutron in lattice QCD with $2 + 1$ flavors of dynamical quarks including quark-connected and -disconnected contributions.
At the same time, we have studied all relevant systematic effects, \ie the contamination by excited states as well as discretization and finite-volume effects.
From direct fits of the form factors to the expressions resulting from B$\chi$PT, we have extracted the electromagnetic radii and magnetic moments of the proton and neutron.
Furthermore, we have computed the Zemach radius of the proton from an extrapolation of our form factors to arbitrarily large $Q^2$-values.
The overall precision of our results for the electromagnetic radii as well as for the Zemach radius is sufficient to make a meaningful contribution to the ongoing debates surrounding these quantities.
In order to fully resolve the still existing tensions, further investigations are clearly required, both on the theoretical and on the experimental side, in particular for the magnetic radius of the proton.

\acknowledgments
This research is partly supported by the Deutsche Forschungsgemeinschaft (DFG, German Research Foundation) through the Cluster of Excellence \enquote{Precision Physics, Fundamental Interactions and Structure of Matter} (PRISMA${}^+$) funded by the DFG within the German Excellence Strategy, and through project HI 2048/1-2.
Calculations for this project were partly performed on the HPC clusters \enquote{Clover} and \enquote{HIMster2} at the Helmholtz Institute Mainz, and \enquote{Mogon 2} at Johannes Gutenberg University Mainz (\url{https://hpc.uni-mainz.de}).
The authors also gratefully acknowledge the support of the John von Neumann Institute for Computing and Gauss Centre for Supercomputing e.V. (\url{https://www.gauss-centre.eu}) for projects CHMZ21, CHMZ36, NUCSTRUCLFL, and GCSNUCL2PT.

\setlength{\bibsep}{0pt}
\bibliography{literature.bib}
\end{document}